\documentclass[prd, preprint, 12pt]{revtex4-1}

\usepackage{amsmath}
\usepackage{amssymb}
\usepackage{setspace}
\usepackage{graphicx}
\usepackage{bbm}
\usepackage{float}
\usepackage{hyperref}

\begin{document}
 
 %

\begin{center}
 {  \large {\bf Space and Time as a Consequence of GRW Quantum Jumps}}

\medskip


{\bf Tejinder P. Singh}

{\it Tata Institute of Fundamental Research,}
{\it Homi Bhabha Road, Mumbai 400005, India}

\end{center}
\bigskip
\setstretch{1.24}


\centerline{\bf ABSTRACT}

\noindent The Ghirardi-Rimini-Weber (GRW) theory of spontaneous collapse offers a possible resolution of the quantum measurement problem. In this theory, the wave function of a particle spontaneously and repeatedly localises to one or the other random position in space, as a consequence of the hypothesised quantum jumps. In between jumps the wave function undergoes the usual Schr\"{o}dinger evolution. In the present paper we suggest that these jumps  take place in Hilbert space, with no reference to physical space, and physical three-dimensional space arises as a consequence of localisation of macroscopic objects in the universe. That is, collapse of the wave-function is responsible for the origin of space. We then suggest that similar jumps take place for a hypothetical time operator in Hilbert space, and classical time as we know it emerges from localisation of this time operator for macroscopic objects. More generally, the jumps are suggested to take place in an operator space-time in Hilbert space, leading to an emergent classical space-time.



\noindent 

\bigskip
\bigskip

\centerline{\textit {\textbf {This article is respectfully dedicated to the memory of Giancarlo Ghirardi}}}

\bigskip



\bigskip
\section{The GRW Theory}
The GRW theory \cite{Ghirardi:86, Ghirardi2:90} makes the following two postulates for dynamics in non-relativistic quantum mechanics:

1. Given the wave function $\psi ({\bf x_1}, {\bf x_2}, ..., {\bf x_N})$ of an $N$ particle quantum system in Hilbert space, the $n$-th particle undergoes  a `spontaneous collapse' to a random spatial position ${\bf x}$ as defined by the following jump operator:
\begin{eqnarray}
{\psi_{t}({\bf x}_{1}, {\bf x}_{2}, \ldots {\bf x}_{N}) \quad
\longrightarrow \quad} 
 \frac{L_{n}({\bf x}) \psi_{t}({\bf x}_{1},
{\bf x}_{2}, \ldots {\bf x}_{N})}{\|L_{n}({\bf x}) \psi_{t}({\bf
x}_{1}, {\bf x}_{2}, \ldots {\bf x}_{N})\|}
\end{eqnarray}

The jump operator $L_{n}({\bf x})$ is a linear operator defined to be the normalised Gaussian:
\begin{equation}
L_{n}({\bf x}) =
\frac{1}{(\pi r_C^2)^{3/4}} e^{- ({\bf
\hat q}_{n} - {\bf x})^2/2r_C^2}
\end{equation}
${\bf \hat q}_{n}$ is the position operator for the $n$-th particle of the system and the random variable ${\bf x}$ is the spatial position to which the jump occurs. $r_C$, the width of the Gaussian, is a new constant of nature.

The probability density for the $n$-th particle to jump to the position
${\bf x}$ is assumed to be given by:
\begin{equation}
p_{n}({\bf x}) \quad \equiv \quad \|L_{n}({\bf x}) \psi_{t}({\bf
x}_{1}, {\bf x}_{2}, \ldots {\bf x}_{N})\|^2
\end{equation}
Also, it is assumed in the GRW theory that the jumps are distributed in time as
a Poissonian process with frequency $\lambda_{\text{\tiny GRW}}$. This is the second
new parameter of the model.

2. Between
two consecutive jumps, the state vector evolves according to the
standard Schr\"odinger equation.

These two postulates together provide a unified description of microscopic and macroscopic dynamics, and an elegant solution to the quantum measurement problem (for reviews see e.g. \cite{Bassi:03,RMP:2012}). 

\section{Physical space as a consequence of the GRW quantum jumps}
If the GRW jumps localise a particle in space, then it is obvious that space is assumed to pre-exist, so that a jump can occur in space. Assuming that the GRW theory is the correct description of modified quantum mechanics, imagine for a moment that no GRW jumps have yet taken place in the universe. There would then be no localised macroscopic objects at all, and every particle will be undergoing Schr\"{o}dinger evolution. If everything in the universe is quantum mechanical, it is not physically meaningful to talk of space. According to the Einstein hole argument \cite{Carlip2001, Singh:2006} in order to assign a physical and operational meaning to space, there must overlie on the space a well-defined classical metric. This classical metric is in turn produced by classical bodies, according to the laws of general relativity. And classical bodies are themselves the result of GRW localisation. Thus it is not reasonable to assume space to exist prior to the GRW quantum jumps. Rather, it seems quite intuitive that space results from GRW collapses taking place all over the universe. Space is that which is between collapsed objects. No collapse, no space. This also helps us understand why the GRW jumps take place in space: it is because space in the first place is created because of these jumps.

We can state this formally, by revising the first postulate of GRW theory. Hilbert space is assumed to be more fundamental than space, and space is assumed to arise as a consequence of the GRW jumps, which actually take place in Hilbert space. We define a set of three new self-adjoint `space operators' ${\bf \hat x}$ which commute with each other and with the ${\bf \hat q}_n$s.. The state of the system is described by the wave function $\psi ({\bf x_1}, {\bf x_2}, ..., {\bf x_N})$, where ${\bf x_n}$ is a set of three degrees of freedom associated with the $n$-th particle, these being real eigenvalues of the newly introduced `space operator' ${\bf \hat x}$ which belongs to the Hilbert space. The state evolves with time according to the following two postulates, which are essentially the same as the GRW postulates, except that one gets rid of classical physical space.

1. Given the wave function $\psi ({\bf x_1}, {\bf x_2}, ..., {\bf x_N})$ of an $N$ particle quantum system in Hilbert space, the $n$-th particle undergoes  a `spontaneous collapse' to a random eigenvalue ${\bf x}$ of ${\bf \hat x}$, as defined by the following jump operator:
\begin{eqnarray}
{\psi_{t}({\bf x}_{1}, {\bf x}_{2}, \ldots {\bf x}_{N}) \quad
\longrightarrow \quad} 
 \frac{L_{n}({\bf x}) \psi_{t}({\bf x}_{1},
{\bf x}_{2}, \ldots {\bf x}_{N})}{\|L_{n}({\bf x}) \psi_{t}({\bf
x}_{1}, {\bf x}_{2}, \ldots {\bf x}_{N})\|}
\end{eqnarray}

The jump operator $L_{n}({\bf x})$ is a linear operator defined to be the normalised Gaussian:
\begin{equation}
L_{n}({\bf x}) =
\frac{1}{(\pi r_C^2)^{3/4}} e^{- ({\bf
\hat q}_{n} - {\bf x})^2/2r_C^2}
\end{equation}
${\bf \hat q}_{n}$ is the position operator for the $n$-th particle of the system and the random variable ${\bf x}$ is the eigenvalue of ${\bf \hat x}$  to which the jump occurs. $r_C$, the width of the Gaussian, is a new constant of nature.

The probability density for the $n$-th particle to jump to the eigenvalue  ${\bf x}$ of ${\bf \hat x}$ is assumed to be given by:
\begin{equation}
p_{n}({\bf x}) \quad \equiv \quad \|L_{n}({\bf x}) \psi_{t}({\bf
x}_{1}, {\bf x}_{2}, \ldots {\bf x}_{N})\|^2
\end{equation}
Also, it is assumed  that the jumps are distributed in time as
a Poissonian process with frequency $\lambda_{\text{\tiny GRW}}$. This is the second
new parameter of the model.

2. Between two consecutive jumps, the state vector evolves according to the standard Schr\"odinger equation.

The particles described by ${\bf \hat q_n}$ `live' in the ${\bf \hat x}$ operator space, and the aforesaid ${\bf x}$ are actually eigenvalues of ${\bf \hat{x}}$. Collapse localises a particle to one of the eigenvalues of ${\bf \hat{x}}$. Macroscopic objects collapse extremely rapidly. Using these eigenvalues as reference points, one interprets the unoccupied eigenvalues around macroscopic objects as the three-dimensional classical physical space we are so familiar with. A quantum mechanical particle which has not undergone collapse still `lives' in the spatial operator space ${\bf \hat{x}}$, but in our conventional formulation of quantum mechanics we work not with operator space, but with classical space, from which the configuration space is constructed. 

\section{GRW quantum jumps in a time operator}
The argument given at the beginning of Section II, namely that it is not meaningful to talk of classical space without classical bodies being present, applies to classical time as well. The Einstein hole argument suggests that in the absence of classical matter fields or material bodies, it is not physically meaningful to assume the existence of a classical space-time manifold. With this in view, we propose that there exists in Hilbert space a self-adjoint `time' operator ${ \hat t}$. In addition we shall assume the existence of a universal classical `scalar time' $s$ whose origin we will discuss in the subsequent section. This time $s$, which keeps track of evolution, we shall refer to as `trace time', for reasons which are clarified in the next section. The wave function will now depend on the extended configuration space $({\bf x_1}, t_1, {\bf x_2}, t_2, ..., {\bf x_N}, t_N)$ where $t_n$ is an eigenvalue of the operator time ${ \hat t}$, and the degrees of freedom $({\bf x}_n, t_n)$ are associated with the `position' operator ${\bf \hat q}_n$ of the $n$-th particle. 
The GRW postulates for collapse in the time operator are stated as follows:

1. Given the wave function $\psi ({\bf x_1}, t_1, {\bf x_2}, t_2, ..., {\bf x_N}, t_N)$ of an $N$ particle quantum system in Hilbert space, the $n$-th particle undergoes  a `spontaneous collapse' to a random eigenvalue ${t}$ of ${ \hat t}$, as defined by the following jump operator:
\begin{eqnarray}
{\psi_{s}({\bf x}_{1}, t_1, {\bf x}_{2}, t_2, \ldots {\bf x}_{N}, t_N) \quad
\longrightarrow \quad} 
 \frac{L_{n}({t}) \psi_{s}({\bf x}_{1}, t_1,
{\bf x}_{2}, t_2,\ldots {\bf x}_{N}, t_N)}{\|L_{n}({t}) \psi_{s}({\bf
x}_{1}, t_1, {\bf x}_{2}, t_2, \ldots {\bf x}_{N}, t_n)\|}
\end{eqnarray}

The jump operator $L_{n}({t})$ is a linear operator defined to be the normalised Gaussian:
\begin{equation}
L_{n}({t}) =
\frac{1}{(\pi t_C)^{1/2}} e^{- (
{\hat t} - {t})^2/2 t _C^2}
\end{equation}
$\hat{\bf q}_{n}$ is the `position' operator for the $n$-th particle of the system and the random variable ${t}$ is the eigenvalue of ${\hat t}$  to which the jump occurs. $t_C$, the width of the Gaussian, is a new constant of nature.

The probability density for the $n$-th particle to jump to the eigenvalue  ${t}$ of ${ \hat t}$ is assumed to be given by:
\begin{equation}
p_{n}({ t}) \quad \equiv \quad \|L_{n}({t}) \psi_{s}({\bf
x}_{1}, t_1, {\bf x}_{2}, t_2, \ldots {\bf x}_{N}, t_N)\|^2
\end{equation}
Also, it is assumed  that the jumps are distributed in trace time $s$ as
a Poissonian process with frequency $\eta_{\text{GRW}}$. This is the fourth
new parameter of the model.

2. Between two consecutive jumps, the state vector evolves according to the following generalised Schr\"odinger equation
\begin{equation}
i\hbar \frac{\partial\psi}{\partial s} = H \psi (s)
\end{equation}
With regard to time, the particles described by ${t_n}$ `live' in the ${ \hat t}$ operator space, and the aforesaid ${t}$ values are actually eigenvalues of ${\hat{t}}$. Collapse localises a particle to one of the eigenvalues of ${ \hat{t}}$. Using these eigenvalues as reference points, one interprets the collection of eigenvalues as the one dimensional classical time we are so familiar with. Time could be said to be that which is between GRW jumps in the operator time. A quantum mechanical particle which has not undergone collapse still `lives' in the time operator space ${ \hat{t}}$, but in our conventional formulation of quantum mechanics we work not with operator time, but with classical time. 
Classical space and time are thus approximations to the operator space and time described by $({\bf \hat{x}}, \hat{t})$, the approximation being caused by GRW quantum jumps.

It seems reasonable to assume that $\eta_{\text{GRW}} = \lambda_{\text{GRW}}$ and that the amplification mechanism works for localisation in the temporal direction, same way that it works for spatial localisation of macroscopic objects.

Our discussion so far has been in the context of non-relativistic quantum mechanics. We next consider how it might be possible to extend it to a relativistic space-time, and try to arrive at classical Lorentz-invariant Minkowski space-time as a consequence of GRW quantum jumps in operator space-time.

\section{GRW jumps in the space-time operator}
We now propose that there exists in the Hilbert space an operator space-time `Minkowski' metric, for which the line-element is
\begin{equation}
ds^2 = {\rm Tr}\; d\hat{s}^2 \equiv {\rm Tr} [c^2\;  d\hat{t}^2 - d {\bf \hat {x}^2 } ] 
\label{ost}
\end{equation}
which also defines the trace time $s$, in terms of the trace of the operator line element. This line-element is invariant under Lorentz transformations of the space-time operator, and allows us to construct a Lorentz invariant classical matrix dynamics. This was shown by us in an earlier work \cite{Lochan-Singh:2011}.  For our present purpose we define a space-time operator $\hat{x}^\mu$ as $x^\mu = (\hat t, {\bf \hat x})$, where all the four operators commute with each other and with the ${\bf \hat q_n}$s. The state  of the system is labelled by eigenvalues of $\hat{x}^{\mu}$, and is hence written as
$\psi(x^\mu_1, x^\mu_2, ..., x^\mu_N)$. Evolution is governed by the trace time $s$ defined above.
We also make the reasonable assumption that the temporal correlation interval $t_C$ is related to $r_C$ by
$t_C = r_C / c$.  Thus we still have only two new parameters, just as in the GRW model. The dynamics is then given by the following variation of the two GRW postulates.

1. Given the wave function $\psi(x^\mu_1, x^\mu_2, ..., x^\mu_N)$ of an $N$ particle quantum system in Hilbert space, the $n$-th particle undergoes  a `spontaneous collapse' to a random eigenvalue $x^{\mu}$ of ${ \hat x}^\mu$, as defined by the following jump operator:
\begin{eqnarray}
{\psi_{s}(x^\mu_1, x^\mu_2, ..., x^\mu_N)\quad
\longrightarrow \quad} 
 \frac{L_{n}({x^\mu}) \psi_{s}(x^\mu_1, x^\mu_2, ..., x^\mu_N)}{\|L_{n}({t}) \psi_{s}(x^\mu_1, x^\mu_2, ..., x^\mu_N)\|}
\end{eqnarray}

The jump operator $L_{n}({x^\mu})$ is a linear operator defined to be the normalised Gaussian:
\begin{equation}
L_{n}(x^\mu) =
\frac{1}{(\pi r_C)^{2}} e^{- ({
{ \hat q}^\mu_n} - {x^\mu})^2/2 t _C^2}
\end{equation}
$\hat{q}^\mu_{n}$ is the position operator for the $n$-th particle of the system and the random variable ${x^\mu}$ is the eigenvalue of ${\hat x}^\mu$  to which the jump occurs. 

The probability density for the $n$-th particle to jump to the eigenvalue  ${x^\mu}$ of ${ \hat x}^\mu$ is assumed to be given by:
\begin{equation}
p_{n}({ x^\mu}) \quad \equiv \quad \|L_{n}({x^\mu}) \psi_{s}(x^\mu_1, x^\mu_2, ..., x^\mu_N)\|^2
\end{equation}
Also, it is assumed  that the jumps are distributed in trace time $s$ as
a Poissonian process with frequency $\eta_{\text{GRW}}=\lambda_{GRW}$. 

2. Between two consecutive jumps, the state vector evolves according to the following generalised Schr\"odinger equation
\begin{equation}
i\hbar \frac{\partial\psi}{\partial s} = H \psi (s)
\label{ose}
\end{equation}
With regard to space-time, the particles described by ${q^\mu_n}$ `live' in the ${ \hat x}^\mu$ operator space, and the aforesaid ${x^\mu}$ values are actually eigenvalues of ${\hat{x}^\mu}$. Collapse localises a particle to one of the eigenvalues of ${ \hat{x}^\mu}$. Using these eigenvalues as reference points, one interprets the collection of eigenvalues as the four dimensional classical spcae-time we are  familiar with. Space-time could be said to be that which is between GRW jumps in the operator space-time. A quantum mechanical particle which has not undergone collapse still `lives' in the space-time operator space ${ \hat{x}^\mu}$, but in our conventional formulation of quantum mechanics we work not with operator space-time, but with classical space-time. It is implicitly assumed that classical proper time coincides with the trace time $s$. 
Classical space and time are thus approximations to the operator space and time described by $({\bf \hat{x}}, \hat{t})$, the approximation being caused by GRW quantum jumps. One could say that these jumps make quantum theory consistent with the Einstein hole argument. Linear quantum theory without these jumps disagrees with the hole argument. Moreover, the recovered space-time possesses Lorentz invariance because the
original operator space-time in (\ref{ost}) is Lorentz invariant. Thus our point of view is that spontaneous collapse is essential for making quantum theory consistent with classical space-time. This gives a motivation for GRW quantum jumps over and above their ad hoc construction with the express motivation of solving the measurement problem. 

After classical space-time has been recovered one can write the Schr\"odinger equation (\ref{ose}) on the standard Minkowski background and transform to the Heisenberg picture, and also develop relativistic quantum field theory in the standard  manner.
While one could take (\ref{ose}) as a starting point, it can also be derived from an underlying matrix dynamics for matter and operator space-time degrees of freedom, as has been done in \cite{Lochan-Singh:2011, Lochan:2012} following the earlier work of Adler and collaborators \cite{Adler:04} on Trace Dynamics.

The following additional remarks serve to clarify the above discussion, and also describe work currently in progress to develop this idea further:

(i) The problem of freezing in time: Just as a macroscopic object spontaneously collapses to a specific position in space and repeated collapses keep it there, spontaneous collapses in time keep it frozen at a specific value of classical time. How then does it evolve in time? One possible solution is to propose that spontaneous collapse takes place not onto space-time points, but to space-time paths. Paths are more fundamental than points. Instead of constructing paths from points, we should construct points from paths. Evolution in time is then a perception - the entire space-time path is in fact pre-given, in the spirit of the principle of least action, which determines the entire path in one go. The mathematical formulation of this proposal is presently being attempted, following the construction of a path-integral for the GRW model \cite{pint:2018}.

(ii) The overall picture: The underlying layer of dynamics is the Hilbert space endowed with the operator space-time metric proposed here. We call this the Extended Hilbert Space (EHS). If every object in this extended Hilbert space undergoes spontaneous localisation, we will have a universe filled with classical objects, and a classical space-time. We may express this as:

Extended Hilbert Space $\rightarrow$ Spontaneous Localisation $\rightarrow$ Classical matter + Classical Spacetime 

However, we know that microscopic objects do not undergo rapid spontaneous localisation; their dynamics must still be described using the EHS. What we do however, is that we carry out the dynamics of the uncollapsed objects, in ordinary Hilbert space, and with reference to classical space-time, as if:

Extended Hilbert Space = Ordinary Hilbert Space + Classical Matter + Classical Spacetime

In so doing, we forget the operator space-time metric (OST), and this omission is at the heart of the quantum measurement problem. We should either stay in EHS (QM + OST) or in the classical universe (CM + CST). Instead we either stay in CM + CST, or we invoke QM + CST. This latter is an approximation; albeit an excellent one for microscopic systems, but the approximation breaks down for macroscopic objects, because spontaneous localisation in the EHS kicks in for them. If we do want to carry out QM on CST, we should actually  approximate QM + OST as

\centerline{QM + OST =  QM + CST + Fluctuations}

\noindent where `Fluctuations' represents the fluctuating aspect of the OST, which must necessarily accompany the mean, the latter being the CST, and which is responsible for spontaneous localisation. It is possible that these fluctuations are the imaginary stochastic component of the metric which has been proposed by Adler \cite{Adler2014}. The reason for the anti-Hermitean nature of the fluctuations remains to be understood. There could also be an intriguing connection with the work of Karolyhazy \cite{Karolyhazi:86} and of Diosi \cite{Diosi:84}, and with the 
Diosi-Penrose model \cite{Diosi:89}. Work is in progress to understand this connection. What is clear though is that spontaneous collapse takes place in EHS, not in ordinary Hilbert space + CST. 

(iii) Emergence of a Lorentz invariant classical space-time: One can consider the classical  line-element $(c^2 dt^2 - d{\bf x}^2)$
to be one of the eigenvalues of the operator line element $( c^2\;  d\hat{t}^2 - d {\bf \hat {x}^2 } )$ and the Lorentz invariance of the latter ensures the Lorentz invariance of the former.

(iv) Emergence of quantum field theory: The underlying EHS framework which we have employed, deals with many particle relativistic quantum mechanics in the EHS. One could well ask as to how this will be generalised to  a quantum field theory? At present however, it is not obvious to us if a field theoretic description is at all required in EHS. The point being that EHS does not have a light cone structure, and does not obey classical special relativity. There is no physical reason for the particles to stay inside the classical light-cone, and it may well be that the quantum field theoretic description is necessitated only for QM on CST. This aspect is at present under investigation.

(v) The role of the Einstein hole argument: According to the Einstein hole argument, one cannot identify classical space-time points without an overlying classical metric. If only quantum matter is present in the universe, the metric cannot be classical, and will undergo fluctuations. Even in a Minkowski space-time, the flat metric will have fluctuations, which will prevent the points of the space-time manifold from being identified. That is why, even in such a situation, one should replace the classical space-time manifold by an alternative structure, e.g. the operator space-time proposed here. The GRW jumps make the matter classical, which in turn no longer has quantum fluctuations in position, and hence produces a classical metric, and in the limit, a Minkowski spacetime without any metric fluctuations, thus permitting a point structure and the manifold.

(vi) Space-time from decoherence as an alternative explanation: Another well-known mechanism for emergence of classical 
space-time from a quantum gravitational regime is decoherence \cite{Schlosshauer:2007} of the different branches of the emergent classical universes, in the semiclassical approximation. This should be considered an in-principle valid explanation, so long as one also assumes the 
validity of the many-world interpretation of quantum mechanics. The model presented in this paper becomes a necessary one though, if one assumes that the GRW quantum jumps are the correct way to solve the measurement problem.

(vii) The rigged Hilbert space: Usually, position operators are not defined in
a (proper) Hilbert space, but in a more general space called
rigged Hilbert space, using Gel'fand triples for its definition, so that bound state (discrete) and continuum spectrum can be treated in a unified setting.  Strictly speaking, our discussion should take place in a rigged Hilbert space; however our analysis has not reached this level of rigour, and furthermore we do not expect the central features of our analysis to depend on this aspect.

(viii) Understanding quantum non-locality: The paradoxical nature of the non-local EPR quantum correlations no longer arises in the extended Hilbert space, as we explain now. Consider an entangled state of two particles, one at the spatial location $x$ in CST, and the other far away, at $y$. Let $O$ be the observable whose eigenvalues $O_{+}$ and $O_{-}$ are correlated for the two particles, so that we write the state-vector for the correlated pair, in CST, as
\begin{equation}
|\psi> = |\psi_x(x,t, O_{+})> \; |\psi_y(y,t,O_{-})> + \; |\psi_x(x,t,O_{-})> \; |\psi_y(y,t,O_{+})> 
\end{equation}
Suppose Alice makes the measurement on particle $x$ at time $t=t_1$ and finds the observable $O$ to have the value $O_{+}$. The state $|\psi>$ then instantaneously collapses to $|\psi_x(x,t, O_{+})> \; |\psi_y(y,t,O_{-})>$, i.e. the state of particle $y$ collapses to the observable value $O_{-}$ at the same instant $t_1$ at which Alice made  the measurement on particle $x$. If 
Bob makes a measurement on particle $y$ at time $t_2$ he will find it in the state $O_{-}$, even if $(y-x)/(t_2-t_1) > c$. This of course is the non-locality puzzle.

Consider now the description of the state in the Extended Hilbert Space. The evolution of the state is tracked by the trace time $s$, which is identical and simultaneous for the two particles. However, because time is an operator, i.e. $\hat{t}$, the two particles are now labelled by {\it different} eigenvalues of operator time, say $t'$ and $t"$. Thus, unlike in the CST case, the entangled state will now be written as
\begin{equation}
|\psi> = |\psi_{x,t'}(x,t',s, O_{+})> \; |\psi_{y,t"}(y,t",s,O_{-})> + \; |\psi_{x,t'}(x,t',s, O_{-})> \; |\psi_{y,t"}(y,t",s,O_{+})> 
\end{equation}
When Alice makes a measurement on particle $x$, at trace time $s$, and finds the observable $O$ to have the value $O_{+}$, the time eigenvalue $t'$ associated with particle $x$ collapses to the time coordinate of Alice, say $t'_{A}$ (the CST time being $t_1$). This measurement instantaneously affects particle $y$, but instantaneously in trace time $s$. The time eigenvalue $t"$ associated with particle $y$ is not affected at all. Moreover, there being no notion of classical distance in the EHS, there is nothing that travels from $x$ to $y$. It is only in CST, that we can talk of $y$ being far from $x$. When Bob makes a measurement on particle $y$, the time eigenvalue $t''$ associated with particle $y$ collapses to the time coordinate of Bob, say $t''_{B}$ (the CST time being $t_2$). Clearly, there is nothing puzzling if $(y-x)/(t_2-t_1) > c$; this has nothing to do with the physics which is taking place in EHS. We realise that notions such as distance, light cone, space-like separation, are imposed on us by CST, and the non-locality puzzle arises only if we refer the measurements to CST, instead of OST.

(ix) The origin of the Born probability rule: It is known that the continuum limit of the GRW model is constructed by assuming the existence of a stochastic noise field, which is added as an anti-Hermitean term to the Schr\"{o}dinger equation. The Born probability rule for collapse follows provided one assumes that norm of the state vector be preserved despite adding an anti-Hermitean term to the Hamiltonian, and if one assumes the absence of superluminal signalling. The condition of norm preservation seems rather ad hoc when seen from the CST. However, in the EHS with an OST, it seems reasonable to assume that the evolution of the state vector given by (\ref{ose}) is actually geodesic evolution in the operator space-time metric and geodesic evolution preserves length of the vector, i.e. its norm. Hence the CST representation of this evolution, in terms of mean and fluctuations, should also be such that the norm is preserved. Born rule is thus seen as a consequence of geodesic motion in a Lorentz invariant operator space-time metric.

(x) Quantum interference in time as evidence for the operator time of EHS? : There are reported results of laboratory experiments which, according to the authors, are strong evidence of quantum interference in time, rather than in space \cite{Gozdz,Czachor}. This would be very hard to understand if time is  a parameter, as in QM  on CST. However, in the EHS, i.e. in QM on OST, a particle has a non-zero amplitude and probability to be simultaneously (i.e. with respect to trace time $s$), at two different times $t'$ and $t"$. Thus interference in time is entirely possible, and we regard the reported experimental evidence for this as strong evidence for operator space-time. Further investigation of this aspect is currently in progress.

An outstanding open question is to understand how gravity emerges from the underlying operator space-time. Nonetheless, it is hoped that the ideas presented here will be of some help in developing a relativistic version of collapse models. The idea that space might result from collapse of he wave function has been proposed earlier by Pearle \cite{PEARLE1}. Diosi and Tilloy have proposed Newtonian gravity as resulting from spontaneous localisation \cite{Diosi:2009, Tilloy:2016,Tilloy:2018}

After this work appeared, there has appeared another paper \cite{MSm} which introduces the important concept of a delocalised time. However, this concept has already been introduced by the present author in the current paper, and in his earlier work
\cite{Singh:2006,Lochan-Singh:2011, Lochan:2012,Singh:2012,Banerjee:2016,Singh:2017,SinghCQ:2017,QTST:2017}.

\bigskip

\noindent {\bf Acknowledgements:} I would like to thank Kinjalk Lochan for comments that significantly improved an earlier version of the manuscript. It is also a pleasure to thank Branislav Nikolic and Antoine Tilloy for helpful correspondence.This work is partially supported by a Mini-Grant from the Foundational Questions Institute.

\vskip 0.3 in


\centerline{\bf REFERENCES}

\bibliography{biblioqmtstorsion}

\noindent\end{document}